\documentclass[preprint,12pt]{elsarticle}

\usepackage{graphics}
\usepackage{threeparttable}
\usepackage{color}
\usepackage{amssymb}
\usepackage{amsthm}

\journal{Chemical Physics}
\begin{document}

\begin{frontmatter}

\title{Facile implementation of integrated tempering sampling method to enhance the sampling over a broad range of temperatures}

\author[label1]{Peng Zhao}
\author[label2]{Li Jiang Yang}
\author[label2]{Yi Qin Gao\corref{cor2}}
\cortext[cor2]{Corresponding author.}
\ead{gaoyq@pku.edu.cn.}
\author[label1]{Zhong-Yuan Lu\corref{cor1}}
\cortext[cor1]{Corresponding author.}
\ead{luzhy@jlu.edu.cn.}

\address[label1]{State Key Laboratory of Theoretical and Computational Chemistry, Institute of Theoretical Chemistry,Jilin University, Changchun 130023, China}
\address[label2]{College of Chemistry and Molecular Engineering,Beijing National Laboratory of Molecular Sciences,Peking University, Beijing 100871, China}

\begin{abstract}
 Integrated tempering sampling (ITS) method is an approach to enhance the sampling over a broad range of energies and temperatures in computer simulations. In this paper, a new version of integrated tempering sampling method is proposed. In the new approach presented here, we obtain parameters such as the set of temperatures and the corresponding weighting factors from canonical average of potential energies. These parameters can be easily obtained without estimating partition functions. We apply this new approach to study the Lennard-Jones fluid, the ALA-PRO peptide and the single polymer chain systems to validate and benchmark the method.
\end{abstract}

\end{frontmatter}

\section{Introduction}
Because the free energy landscape of typical macromolecular system is rough and complicated with plenty of minima and barriers, it is difficult to search global free energy minimum using conventional molecule dynamics (MD) or Monte Carlo (MC) simulations. In the last decades, a variety of methods have been developed to achieve an extensive sampling of configurational space. These methods include umbrella sampling~\cite{valleau1977umbre,laio2002metad}, replica exchange method(REM)~\cite{sugita1999remd,berg1991mul}, multicanonical simulation~\cite{okabe2001remc,berg1992mul}, metadynamics~\cite{valleau1991umbre}, simulated tempering~\cite{lyubartsev1992st,john2010st}, essential dynamic sampling~\cite{berendsen1996essd}, Wang-Landau algorithm~\cite{wang2001landau,wang2001prl}, temperature accelerated sampling~\cite{okamoto2004gen}, and so on. Many of these methods are based on generalized ensemble~\cite{vander2006tamd} in which each configuration is weighted by a non-Boltzmann probability factor, and thus a random walk in energy space could be achieved, \emph{i.e.} via a multicanonical method. These generalized ensemble methods have been extensively applied to the studies of, for example, spin glass~\cite{berg1992mulspin} and protein folding~\cite{hansmann1999mulpro,Bruce1996mulcon}. However, the non-Boltzmann probability factor is usually unknown and should be determined by iteration processes. These iterations are non-trivial and even difficult for complex systems, therefore some methods are proposed to accelerate the convergence of iteration processes~\cite{bartels1998mulcon,kumar1996multcon}.

 \par
 Recently, an integrated tempering sampling (ITS) method for enhancing the sampling in energy and configuration space was proposed~\cite{gaoyq2008its1,gaoyq2009comp,gaoyq2008its2}. This method is based on a generalized (non-Boltzmann) ensemble which allows an enhanced sampling in a desired broad energy and temperature range. In this generalized ensemble, the probability of a configuration of the system under study is proportional to a summation of Boltzmann factors at a set of temperatures, with each Boltzmann factor carrying a weighting factor. These weighting factors can be determined by the condition that each term in the summation contributes a predefined fraction.

\par
In the original ITS method, the weighting factors can be estimated in an iterative way, which may be time-consuming for large systems. In this study, we follow the line of ITS method and derive the expression of weighting factors through optimizing the energy distribution in the simulations. The values of weighting factors only depend on the average potential energy of the system, which do not have to be very accurate and can be easily calculated by conventional MD or MC simulations. This process avoids iteration, so the weighting factors can be determined easily and quickly. Moreover, the temperature distribution of an ITS simulation is very important. A broad energy distribution cannot be generated unless a proper temperature range is chosen. Here we also propose an easy-to-use way to generate temperature distribution that can ensure a reasonable energy distribution.

\par
This paper is organized as follows. In section~\ref{sec:method}, the theory and computational scheme will be described in detail. In section~\ref{sec:application}, we apply the method to the studies of Lennard-Jones fluid, a small peptide and single polymer chain to validate and benchmark the method. Conclusion is drawn in section~\ref{sec:conclusion}.

\section{Method}\label{sec:method}
\subsection{Generalized ensemble}
ITS method is based on the generalized ensemble to get a distribution covering a broad range of energies. We define the generalized distribution function $W(r)$ as a summation of a set of Boltzmann factors at different temperatures $T_k$:
\begin{equation}
\label{equ:expwr}
W(r) = \sum\limits_k {{n_k}{e^{ {-\beta_k}U(r)}}} \qquad k=1,2,\ldots,N\,.
\end{equation}
In Eq.~(\ref{equ:expwr}), $\beta_k = {1}/{k_BT_k}$, $k_B$ is Boltzmann constant. In this study, we assume that all the terms in the summation are ranked as temperature increases. The probability to find a configuration with potential energy $U$ is proportional to $W(r)$.
Eq.~(\ref{equ:expwr}) shows that the generalized ensemble is closely associated with the canonical ensembles at different temperatures. The properties of the generalized ensemble can be calculated from those canonical ensembles. For example, the partition function is:
\begin{equation}
\label{equ:equqw}
{Q_W} = \int {W(r)dr = } \int {\sum\limits_k {{n_k}{e^{ -{\beta _k}U(r)}}} dr = \sum\limits_k {{n_k}{Q_k}} }\,.
\end{equation}
In Eq.~(\ref{equ:equqw}), $Q_W$ is the partition function of the generalized ensemble and $Q_k$ is the partition function of the canonical ensemble at temperature $T_k$. The ensemble average of thermodynamic quantity $\langle A\rangle_W$ is
\begin{equation}
\label{equ:equaw}
{\left\langle A \right\rangle _W} = \frac{{\int {A(r)W(r)dr} }}{{\int {W(r)} dr}} = \frac{{\int {A(r)\sum\limits_k {{n_k}{e^{ - {\beta _k}U(r)}}dr} } }}{{\int {\sum\limits_k {{n_k}{e^{ - {\beta _k}U(r)}}dr} } }} = \frac{{\sum\limits_k {{n_k}{Q_k}{{\left\langle A \right\rangle }_k}} }}{{\sum\limits_k {{n_k}{Q_k}} }}\,.
\end{equation}
Here, $A$ is a thermodynamic quantity, $\left\langle A \right\rangle _W$ denotes generalized ensemble average of $A$, $\left\langle A \right\rangle _k$ denotes canonical ensemble average. The potential energy probability density of the generalized ensemble $P_W(U)$ is
\begin{equation}
\label{equ:equpw}
{P_W}(U) = \frac{{n(U)W(r)}}{{\int {W(r)} dr}} = \frac{{\sum\limits_k {{n_k}{Q_k}{P_k}(U)} }}{{\sum\limits_k {{n_k}{Q_k}} }}\,,
\end{equation}
 in which $n(U)$ is the density of states, and $P_k(U)$ is the potential energy probability density of the canonical ensemble at temperature $T_k$. In a special case, if  ${n_k} = \frac{c}{Q_k}$ ($c$ is a nonzero constant), Eq.~(\ref{equ:equpw}) becomes
\begin{equation}
\label{equ:simpw}
{P_W}(U) = \frac{{\sum\limits_k {{n_k}{Q_k}{P_k}(U)} }}{{\sum\limits_k {{n_k}{Q_k}} }} = \frac{1}{N}\sum\limits_k {{P_k}(U)}\,.
\end{equation}
Importantly, the properties of any canonical ensemble whose temperature is in the desired range, \emph{i.e.} $T_j\in[T_1,T_N]$ can be calculated by a reweighting scheme from the generalized ensemble by Eq.~(\ref{equ:reweight}) and Eq.~(\ref{equ:prewei}):
\begin{equation}
\label{equ:reweight}
{\left\langle A \right\rangle _{\beta_j} } = \frac{{\int {A(r)} {e^{ - \beta_j U(r)}}dr}}{{\int {{e^{ - \beta_j U(r)}}dr} }} = \frac{{\int {\frac{{A(r){e^{ - \beta_j U(r)}}}}{{W(r)}}W(r)dr} }}{{\int {\frac{{{e^{ - \beta_j U(r)}}}}{{W(r)}}W(r)dr} }} = \frac{{{{\left\langle {\frac{{A(r){e^{ - \beta_j U(r)}}}}{{W(r)}}} \right\rangle }_W}}}{{{{\left\langle {\frac{{{e^{ - \beta_j U(r)}}}}{{W(r)}}} \right\rangle }_W}}}\,,
\end{equation}
\begin{equation}
\label{equ:prewei}
P_{\beta_j}(U) = \frac{n(U) e^{-\beta_j U}}{Q_{\beta_j}} = \frac{ e^{-\beta_j U}}{\sum \limits_k n_k e^{-\beta_k U}} \frac{1}{ \langle \frac{e^{-\beta_j U(r)}} {W(r)} \rangle _W} P_W(U)\,.
\end{equation}
In Eq.~(\ref{equ:prewei}), $P_{\beta_j}(U)$ denotes the potential energy probability density at inverse temperature $\beta_j$ and $Q_{\beta_j}$ denotes the partition function of canonical ensemble at inverse temperature $\beta_j$.

\par
In ITS simulation, the generalized distribution function of Eq.~(\ref{equ:expwr}) can be obtained by running a simulation with a modified potential $U'(r)$ at desired temperature $T$. $U(r)$ is defined through
\begin{equation}
\label{equ:defuprime}
e^{-\beta U'(r)} = W(r)= \sum\limits_k {{n_k}{e^{ {-\beta _k}U(r)}}}\,,
\end{equation}
and can be simply written as:
\begin{equation}
\label{equ:expuprime}
U'(r) = - \frac{1}{\beta}\ln{\sum\limits_k {{n_k}e^{{-\beta_k}U(r)}}}\,.
\end{equation}
The biased force $F_b$ that is used in the Newtonian equations of motion with the modified potential $U'(r)$ becomes
\begin{equation}
\label{equ:expbf}
F_b = - \frac{\partial{U'(r)}}{\partial{r}} = -\frac{\partial{U'(r)}}{\partial{U(r)}}\frac{\partial{U(r)}}{\partial{r}} =\frac{\sum\limits_k{n_k \beta_k e^{-\beta_kU(r)}}}{\beta\sum\limits_k{n_ke^{-\beta_kU(r)}}}{F} .
\end{equation}
In Eq.~(\ref{equ:expbf}), $F$ is the force calculated using original potential function of the system under study. To implement this ITS method in an MD software package, we only need to modify the integrator, which calculates the biased force by Eq.~(\ref{equ:expbf}), leaving other software codes such as subroutines for force calculation unchanged. Therefore, the ITS method supplies an easy and efficient way to scan a larger span of energy distribution.

\subsection{How to determine $n_k$ and $\beta _k$}
 The key issue in ITS method is how to determine the weighting factors $n_k$. In original ITS method~\cite{gaoyq2008its2}, to calculate $n_k$, $m_k$ is defined as
 \begin{equation}
 \label{equ:defmk}
 m_k=\begin{array}{ll}
 1 &\qquad k=1\\
 \frac{n_k}{n_{k-1}} &\qquad k>1 \end{array}\,,
 \end{equation}
 so $n_k$ can be obtained by the product of $m_k$,
 \begin{equation}
 \label{equ:nkmk}
 n_k = n_1\prod\limits_{j=1}^k{m_j}\,,
 \end{equation}
 and $P_k^{con}$ is defined as product of $n_k$ and $Q_k$,
 \begin{equation}
 \label{equ:defgk}
 P_k^{con}=n_k Q_k=n_k \int{e^{-\beta_k U(r) }dr}\,.
 \end{equation}
 In practice, a set of initial guess of $m_k$ is made, then short ITS simulations are performed and $m_k$ are updated in an iterative way to make $P_k^{con}$ of adjacent temperatures equal. Values of $n_k$ are determined by Eq.~(\ref{equ:nkmk}) and the target values of $n_k$ are simply $\frac{c}{Q_k}$ ($c$ is a nonzero constant).

\par
\begin{figure}[!htbp]
\begin{center}
\includegraphics[angle=0,width=0.8\textwidth]{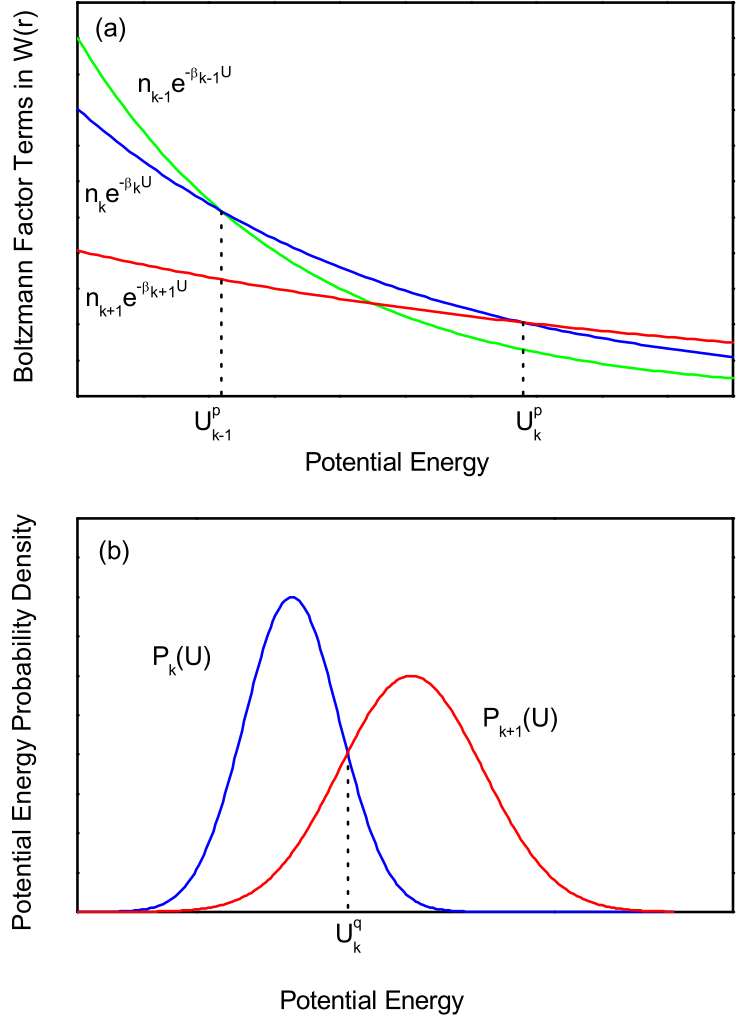}
\end{center}
\caption{Schematic illustration of $U_k^p$ and $U_k^q$. (a) $U_k^p$ is defined as the energy at which the values of two adjacent terms in $W(r)$ are equal. In the range of $U_{k-1}^p<U<U_{k}^p$, the $k$-th term in $W(r)$ is the largest one. (b) $U_k^q$ is defined as the energy at which the potential energy probability density functions of the canonical ensembles at temperature $T_k$ and at temperature $T_{k+1}$ are equal.}
\label{figure:upuq}
\end{figure}

In this study, we propose an alternative way to get the values of $n_k$ quickly, easily and without an iteration. First, we define the energy $U_k^p$ (shown in Fig.~\ref{figure:upuq} (a)), at which the values of two adjacent terms in $W(r)$ are equal. It gives
\begin{equation}
\label{equ:defup}
{n_k}{e^{  {-\beta _k}U_k^p}} = {n_{k+1}}{e^{  {-\beta _{k + 1}}U_k^p}}\,.
\end{equation}
We can easily obtain the expression of $U_k^p$ as
\begin{equation}
\label{equ:expup}
U_k^p = \frac{{\ln {n_k} - \ln {n_{k + 1}}}}{{{\beta _k} - {\beta _{k + 1}}}}\,.
\end{equation}
As mentioned before, terms in $W(r)$ are ranked as temperature increases. According to mathematical property of exponential function, $U_k^p$ increases with increasing temperature:
\begin{equation}
\label{equ:expup2}
U_1^p < U_2^p <  \ldots <U_k^p <  \ldots  < U_{N-1}^p\,.
\end{equation}
This sequence divides the energy into $N$ ranges. Provided that energy $U$ is in the range $U_{k-1}^p < U < U_{k}^p$, the $k$-th term in $W(r)$ is the largest one (as illustrated in Fig.~\ref{figure:upuq} (a)):
\begin{equation}
\label{equ:expup3}
{n_1}{e^{  {-\beta _1}U}} < {n_2}{e^{  {-\beta _2}U}} <  \ldots  < {n_k}{e^{  {-\beta _k}U}} >  \ldots  > {n_N}{e^{  {-\beta _N}U}}\,.
\end{equation}
If we define weighting functions by
\begin{equation}
\label{equ:expup4}
f_k^W(U) = \frac{n_ke^{-\beta_kU}}{\sum\limits_m n_me^{-\beta_mU}}\,,
\end{equation}
there is a maximum of weighting function $f_k^W$ in the range $U_{k-1}^p < U < U_{k}^p$. The value of $f_k^W$ normally decays rapidly as energy $U$ varies. In other ranges, the value of $f_k^W$ could  be rather small even negligible. This property indicates that in the range $U_{k-1}^p < U < U_{k}^p$, the value of $W(r)$ is dominated by its $k$-th term, and the property of generalized ensemble resembles the canonical ensemble at temperature $T_k$.

\par
We then define energy $U_k^q$ (shown in Fig.~\ref{figure:upuq} (b)) meeting the condition that the potential energy probability density function of the canonical ensemble at temperature $T_k$ is equal to that of the canonical ensemble at temperature $T_{k+1}$, that is, ideally we have
\begin{equation}
\label{equ:defuq}
P_k(U_k^q) = P_{k+1}(U_k^q)\,.
\end{equation}
Eq.~(\ref{equ:defuq}) can be written as:
\begin{equation}
\frac{{n(U_k^q){e^{ {-\beta _k}U_k^q}}}}{{{Q_k}}} = \frac{{n(U_k^q){e^{ {-\beta _{k + 1}}U_k^q}}}}{{{Q_{k + 1}}}}\,.
\end{equation}
Then, we can get the expression of $U_k^q$ as
\begin{equation}
\label{equ:expuq}
U_k^q = \frac{{\ln {Q_{k + 1}} - \ln {Q_k}}}{{{\beta _k} - {\beta _{k + 1}}}}\,.
\end{equation}
Provided that the potential energy average of the system will increase as the temperature increases, $U_k^q$ also increases as temperature increases, \emph{i.e}.
$U_1^q < U_2^q <  \ldots  < U_k^q <  \ldots  < U_{N-1}^q$.
And similarly in the range $U_{k-1}^q < U < U_{k}^q$ , there is a maximum for function $P_k(U)$.

\par
To optimize the energy distribution generated in ITS simulation, when $W(r)$ is dominated by the $k$-th term in the range $U_{k-1}^p < U < U_{k}^p$, the maximum of the potential energy probability density function should be in the same range, that is
\begin{equation}
\label{equ:optcon}
U_k^p = U_k^q\,.
\end{equation}
If we substitute Eq.~(\ref{equ:expup}) and Eq.~(\ref{equ:optcon}) into Eq.~(\ref{equ:expuq}), we can conclude that
\begin{equation}
\label{equ:simnk}
{n_k} = \frac{c}{{{Q_k}}}\,.
\end{equation}
Eq.~(\ref{equ:simnk}) is consistent with the result reported in original ITS method, and it indicates that the optimizing condition we present here is essentially identical to the way proposed in Ref.~\cite{gaoyq2008its2}.
If we only substitute Eq.~(\ref{equ:expup}) into Eq.~(\ref{equ:optcon}), we obtain the recursive relation of $n_k$:
\begin{equation}
\label{equ:expnk}
\ln {n_k} - \ln {n_{k + 1}} = U_k^q({\beta _k} - {\beta _{k + 1}})\,.
\end{equation}
In Eq.~(\ref{equ:expnk}), $n_1$ can be simply set to 1 and $U_k^q$ can be estimated in the following way
\begin{equation}
\label{equ:appuq}
U_k^q = \frac{{\ln {Q_{k + 1}} - \ln {Q_k}}}{{{\beta _{k}} - {\beta _{k + 1}}}} \approx  - \frac{1}{2}(\frac{{\partial \ln {Q_k}}}{{\partial {\beta _k}}} + \frac{{\partial \ln {Q_{k + 1}}}}{{\partial {\beta _{k + 1}}}}) = \frac{1}{2}({\left\langle U \right\rangle _k} + {\left\langle U \right\rangle _{k + 1}})\,.
\end{equation}
In Eq.~(\ref{equ:appuq}), the slope of a secant line is approximated by average of the slopes of tangent lines at two line terminals. The potential energy averages can be evaluated through conventional MD simulations. According to Eq.~(\ref{equ:expnk}) and Eq.~(\ref{equ:appuq}), we can easily determine the values of $n_k$ one by one without estimating the partition functions.

\par
The temperature distribution is crucial to the energy distribution generated in ITS simulation. It seriously affects the efficiency of ITS method. Here we also propose an easy way to determine a reasonable temperature distribution, which can actually be determined by the requirement that the ratio between energy probability density functions at two adjacent temperatures is a constant $t$ when the energy is equal to the potential energy average at the lower temperature,
\begin{equation}
\label{equ:defbk}
\frac{{{P_k}({{\left\langle U \right\rangle }_k})}}{{{P_{k + 1}}({{\left\langle U \right\rangle }_k})}} = t\,.
\end{equation}
In Eq.~(\ref{equ:defbk}), the parameter $t$ is named overlap factor, which is related to the space between two adjacent temperatures and the total number of temperatures in the desired temperature range. Eq.~(\ref{equ:defbk}) can be rewritten as
\begin{equation}
\frac{\frac{n({\left\langle U \right\rangle }_k) e^{-\beta_k{\left\langle U \right\rangle }_k}}{Q_k}} {\frac{n({\left\langle U \right\rangle }_{k}) e^{-\beta_{k+1}{\left\langle U \right\rangle }_k}}{Q_{k+1}}}= t\,.
\end{equation}
Through simple derivation, one can get the recursive relation of inverse temperature,
\begin{equation}
\label{equ:expbk}
{\beta _k} - {\beta _{k + 1}} = \frac{{\ln t}}{{U_k^q - {{\left\langle U \right\rangle }_k}}}\,.
\end{equation}
Eq.~(\ref{equ:expbk}) contains only one adjustable parameter $t$. Once the overlap factor $t$ is determined, the temperature distribution will be completely determined.

\par
Because the idea of ITS method is quite similar to that of replica exchange method~\cite{sugita1999remd,okabe2001remc}, we then choose the value of overlap factor $t$ by comparing ITS to REM. REM is based on simultaneous simulations of multiple replicas of the same system at different temperatures. At regular intervals, $N$ independent simulations are allowed to switch temperatures with each other with the acceptance ratio defined in Eq.~(\ref{equ:remdex}). In this way, it is possible for low temperature replicas to gradually migrate up to higher temperatures and back again.
\begin{equation}
\label{equ:remdex}
P_{acc}(U_k,\beta_k \leftrightarrow U_{k+1},\beta_{k+1}) = \min\{1,e^{(\beta_{k+1} - \beta_{k})(U_{k+1} -U_{k})} \}\,.
\end{equation}
In Eq.~(\ref{equ:remdex}), $U_k$ and $U_{k+1}$ are potential energies at temperatures $T_k$ and $T_{k+1}$, respectively. For efficient REM simulations, the choice of temperatures should guarantee sufficient overlap between all adjacent pairs over the entire temperature range and give the same mean acceptance ratio between those adjacent pairs. Various approaches to optimize the temperature distribution of REM simulations had been proposed. Sanbonmatsu and Garc\'{\i}a performed short simulations at a few temperatures, then fitted average energies with polynomial and determined the temperature distribution by solving Eq.~(\ref{equ:remdex}) in an iterative way~\cite{sanbonmatsu2001remd}. de Pablo and coworkers presented a similar approach and demonstrated that under the assumption that the energy probability density function is Gaussian, the relation between acceptance ratio $P_{acc}$ and the overlap of energy probability density function at two adjacent temperatures is system independent~\cite{pablo2004opt}.

\par
For ITS method, substituting Eq.~(\ref{equ:appuq}) into Eq.~(\ref{equ:expbk}), we get:
\begin{equation}
\label{equ:simbk}
{\beta _k} - {\beta _{k + 1}} = \frac{{2\ln t}}{{ {\left\langle U \right\rangle }_{k+1} - {\left\langle U \right\rangle }_k }}\,,
\end{equation}
so,
\begin{equation}
\label{equ:itsex}
e^{(\beta_{k+1} - \beta_{k})( {\left\langle U \right\rangle }_{k+1} - {{\left\langle U \right\rangle }_k}) } = t^{-2}\,.
\end{equation}
The left side of Eq.~(\ref{equ:itsex}) is the mean acceptance ratio in REM simulations. Suppose that if a set of temperatures could give a reasonable acceptance ratio in REM simulations, there should be enough overlap between adjacent temperatures, and this set of temperatures would also work in ITS simulation. Thus, giving the left side of Eq.~(\ref{equ:itsex}) a proper value in the range of $0\sim 1$ will determine the value of overlap factor $t$. Then the complete temperature distribution  is determined further by using Eq.~(\ref{equ:expbk}).

\par
\subsection{Computational procedure}
We propose a new computational procedure of ITS simulation.
\begin{enumerate}
\item Determine the desired temperature range.
\item Choose a set of temperatures in the desired range and generate short replica exchange simulation trajectories to calculate the potential energy averages of the system at those temperatures. For simple systems, conventional MD or MC simulations can also be used. Then the relation between potential energy average and temperature is obtained by interpolation.
\item \label{item:param}Determine the ITS temperature distribution and the corresponding weighting factors $n_k$ through Eq.~(\ref{equ:expnk}), Eq.~(\ref{equ:appuq}) and Eq.~(\ref{equ:expbk}).
\item Use the parameters generated in step~\ref{item:param} to perform ITS simulation, which is essentially a conventional MD simulation using biased force calculated by Eq.~(\ref{equ:expbf}).
\item After ITS simulation, the canonical ensemble properties can be calculated by Eq.~(\ref{equ:reweight}) and Eq.~(\ref{equ:prewei}).
\end{enumerate}

\par
\section{Applications}\label{sec:application}
\subsection{Lennard-Jones fluid}
Lennard-Jones (LJ) fluid is a widely used benchmark system~\cite{ko1993lj}. To test the validity of this ITS method, we consider the LJ fluid system reported in Ref.~\cite{js1993lj} and compare our results with literature data. The LJ potential is
 \begin{equation}
 \label{equ:lj}
 {U_{LJ}}(r) = 4\varepsilon [{(\frac{\sigma }{r})^{12}} - {(\frac{\sigma }{r})^6}].
 \end{equation}

The system contains $864$ particles. In our simulations for LJ particles, conventional reduced units are used. The number density is therefore $\rho=0.8$ and the cutoff distance is $r_c=4.0$. Integration timestep of $0.001$ is used.  Long range correction $U_{tail}$~\cite{allen1989bible} is applied with
\begin{equation}
{U_{tail}} = \frac{8}{9}\pi \rho [{(\frac{\sigma }{{{r_c}}})^9} - 3{(\frac{\sigma }{{{r_c}}})^3}]\,.
\end{equation}

\par
First, we perform a set of conventional (canonical) MD simulations at different temperatures to obtain the potential energy versus temperature curve. Because our purpose is to test the validity of our newly-proposed ITS procedure, we try to obtain the potential energy curve as accurate as possible. So $15$ temperatures are chosen in the range of $1.4\sim 2.0$ for this purpose. At each temperature, we run $1\times10^6$ steps canonical ensemble simulation to calculate the potential energy average after equilibrium. Linear interpolation is applied to obtain the potential energy average in the desired temperature range. Then the temperature distribution $\beta_k$ is obtained by solving Eq.~(\ref{equ:expbk}). The overlap factor $t$ is set to $e^{0.5}$, which generates $9$ temperatures in the range of $1.4\sim 1.91$. The corresponding weighting factors $n_k$ are determined using Eq.~(\ref{equ:expnk}). After that, We perform $1\times10^7$ steps ITS simulation and calculate the canonical average of potential energy at three different temperatures through reweighting scheme using Eq.~(\ref{equ:reweight}). The results are shown in TABLE~\ref{table:ljpot}. The potential energies per particle at different temperatures are in good agreement with the literature data~\cite{js1993lj}. These results indicate that our newly proposed ITS method is validated.

\par
\begin{figure}[!htbp]
\begin{center}
\includegraphics[angle=0,width=1.0\textwidth]{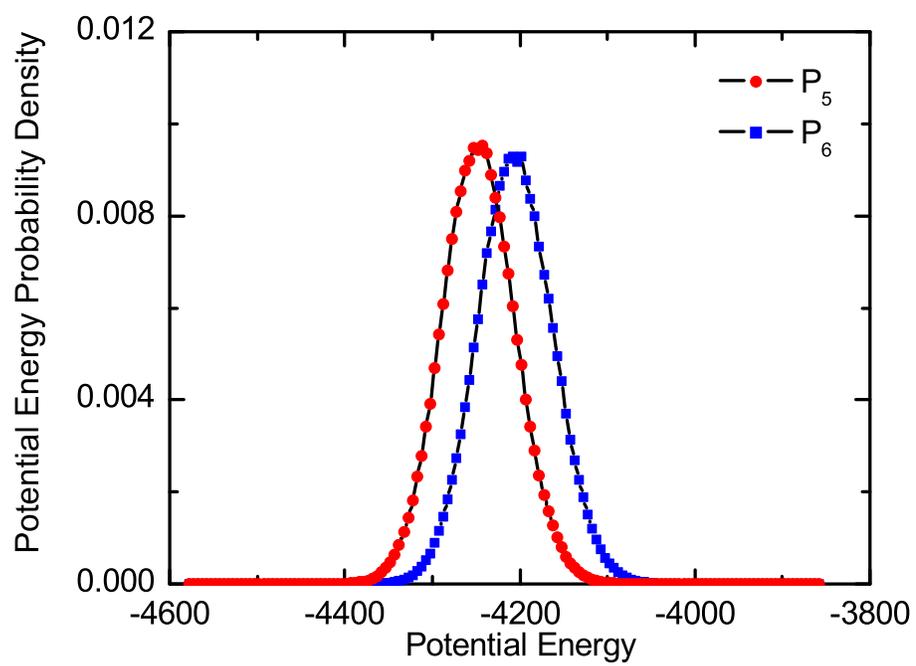}
\end{center}
\caption{The potential energy probability densities of canonical ensembles at temperature $T_5$ (blue square) and $T_6$ (red circle).}
\label{figure:ljpotdis}
\end{figure}

We also calculate the potential energy probability densities of canonical ensembles at $T_5$ and $T_6$ through reweighting scheme using Eq.~(\ref{equ:prewei}). These two temperatures are chosen because they are in the middle of the temperature range between $1.4\sim 1.91$. The results are shown in Fig.~\ref{figure:ljpotdis}. We can see that the two curves overlap largely. Moreover, the value of $U_5^q$ read from Fig.~\ref{figure:ljpotdis} (the cross point of the two curves) is $-4227.7$, in agreement with the value estimated by Eq.~(\ref{equ:appuq}), which is used in the ITS simulation, $-4225.2$. Therefore, our method can ensure sufficient energy distribution overlap between adjacent temperatures, and the approximation method used in Eq.~(\ref{equ:appuq}) yields a good accuracy.

\subsection{ALA-PRO peptide in implicit solvent}
We then apply this ITS method to study the trans/cis transition of ALA-PRO peptide. Real units are used in the following. For comparison, we also use replica exchange molecular dynamics (REMD) and conventional MD to study the conformational transition of this peptide.
\par
\begin{figure}[!htbp]
\begin{center}
\includegraphics[angle=0,width=1.0\textwidth]{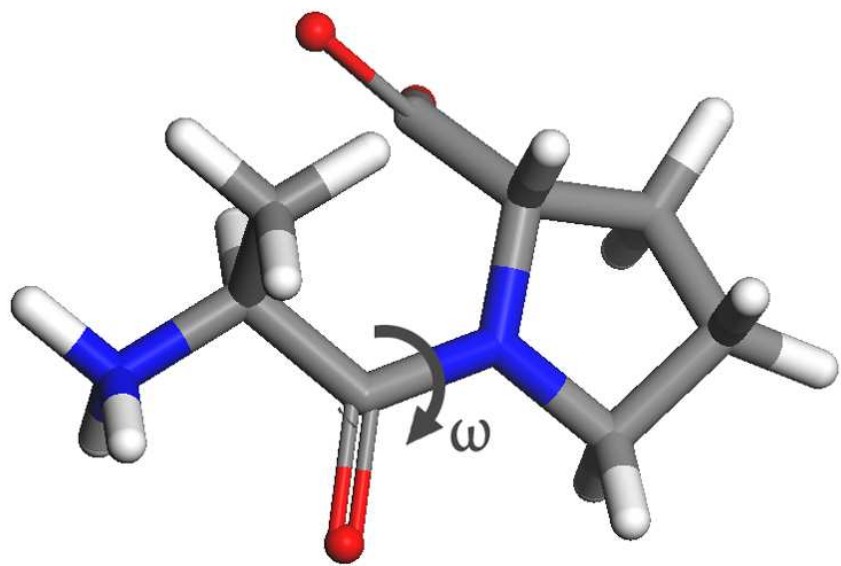}
\end{center}
\caption{The structure of ALA-PRO peptide. The dihedral angle $\omega$ is defined as the reaction coordinate. }\label{figure:ala-pro}
\end{figure}

The structure of ALA-PRO peptide is shown in Fig.~\ref{figure:ala-pro}. The dihedral angle $\omega$ indicated in Fig.~\ref{figure:ala-pro} can be defined as the reaction coordinate of trans/cis transition of this peptide. In our simulations, we use modified GROMACS $4.5.5$ package~\cite{hess2008gromacs} and AMBER $99$sb force field~\cite{sorin2005amber99sb}. Generalized Born solvent accessible surface area (GBSA) implicit solvent model~\cite{tsui2000gbsa} is adopted. The LINCS~\cite{hess1997lincs} algorithm is used to constrain all bonds containing hydrogen atom. In all simulations, the integration timestep is set as $1$ fs. We perform $1$ $\mu$s simulations using ITS, REMD and conventional MD methods, respectively. In REMD simulation, eight replicas with temperatures at $283$, $335$, 400, $478$, $564$, $680$, $805$, and $905$ K are used, which result in an exchange acceptance ratio of roughly $50\%$. The exchange attempt frequency is $1$ ps$^{-1}$ in REMD simulation. For ITS, REMD and conventional MD simulations, the same initial structure is taken.

\par
\begin{figure}[!htbp]
\begin{center}
\includegraphics[angle=0,width=0.8\textwidth]{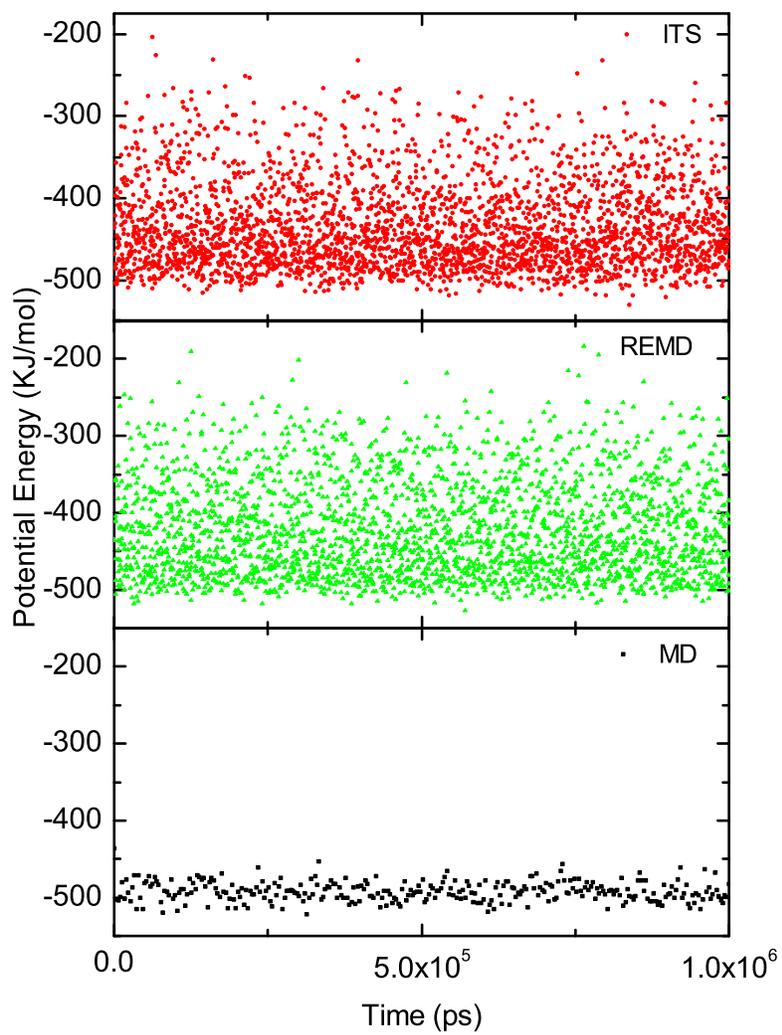}
\end{center}
\caption{Potential energies visited in ITS, REMD and conventional MD simulations. (a) ITS (green triangle); (b) REMD ( red circle); (c) conventional MD (black square).}\label{figure:potener}
\end{figure}

To test the robustness of ITS method, the potential energy average curve is obtained from the first $10$ ps REMD simulations (averaging over 10 frames), which is apparently quite approximate to estimate $n_k$. In our ITS simulation, the overlap factor $t$ is set to $e^{0.05}$, which generates 22 temperatures in the range from $283.0$ to $948.82$ K. The potential energy averages calculated from 10 ps, $1$ $\mu$s REMD simulations and ITS simulation are shown in TABLE~\ref{table:enerpot}. It is clear that the potential energy averages calculated from $10$ ps trajectory of REMD simulation are not accurate enough, whereas potential energy averages calculated from $1~\mu$s REMD and ITS simulation are in good agreement. 
The visited potential energies in ITS, REMD, and conventional MD simulations are shown in Fig.~\ref{figure:potener}. ITS method can effectively explore a broad range of potential energy as REMD method, whereas the conventional MD can only explore a limited potential energy range. An important thing is, although the $n_k$ values are obtained from potential energy averages without enough accuracy (i.e., only from 10 ps REMD trajectories), the ITS method is still quite efficient on exploring the configuration space. It implies that for estimating $n_k$, several short simulations at different temperatures are enough. We denote $\langle U \rangle _k $ and $U_k^{q}$ obtained from short simulations as $\langle U \rangle _k ^{s}$ and $U_k^{qs}$, the true values of $\langle U \rangle _k $ and $U_k^{q}$  are denoted as $\langle U \rangle _k ^{t} $ and $U_k^{qt}$, respectively. The optimizing condition of potential energy distribution is that when $W(r)$ is dominated by the $k$-th term in the range $U_{k-1}^p < U < U_{k}^p$, the maximum of the potential energy probability density should be in the same range. In fact, this optimizing condition can be achieved when the following condition is satisfied:
\begin{equation}
\label{equ:potlim}
U_{k} ^{qs}<\langle U \rangle _{k+1} ^{t} <U_{k+1} ^{qs} (k=1,2,\cdots,N-2) \,.
\end{equation}
Thus, the potential energy averages obtained from short simulations can vary in a quite wide range and are not necessary to be that accurate.

\par
\begin{figure}[!htbp]
\begin{center}
\includegraphics[angle=0,width=1.0\textwidth]{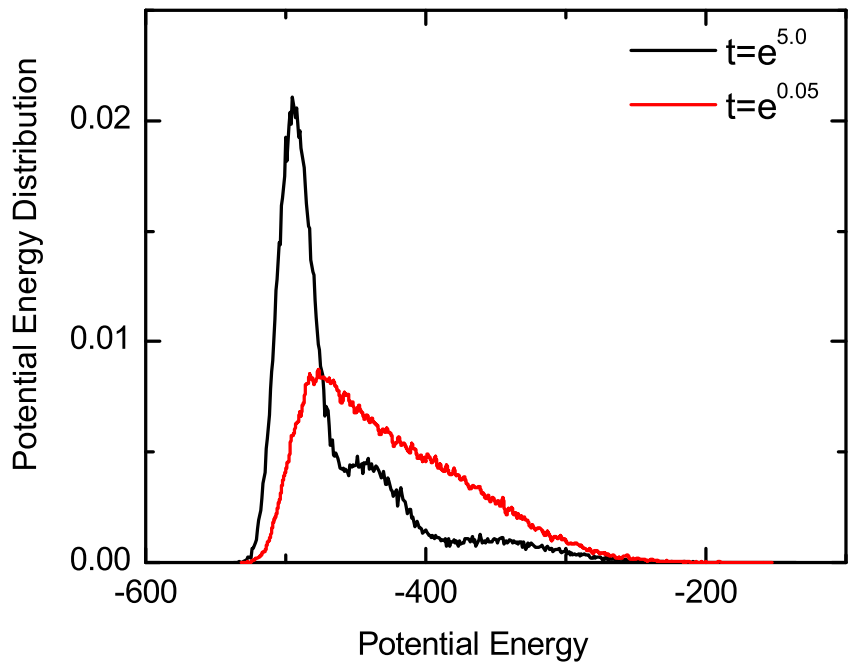}
\end{center}
\caption{Comparison of potential energy distribution generated by $t=e^{5.0}$ and $t=e^{0.05}$.}\label{figure:potdist}
\end{figure}

To clarify the influence of overlap factor $t$ on ITS simulations, we also try different values of overlap factor $t$ for this dipeptide system. The standard deviation of potential energy, $\sigma_d$, is employed to characterize the width of the energy range visited in the ITS simulations:
\begin{equation}
\label{equ:ermsd}
\sigma_d = \sqrt{\langle U^2 \rangle - {\langle U \rangle}^2 }\,.
\end{equation}
The results are shown in TABLE~\ref{table:t}. When $t$ is set to $e^{29.8}$, the overlap between adjacent canonical ensembles is very small, thus the energies of the system under study are trapped in a narrow range. As $t$ decreases, the system can visit a broad range of energies. Fig.~\ref{figure:potdist} shows the potential energy distribution generated by $t=e^{5.0}$ (corresponds to $3$ temperatures) and $t=e^{0.05}$ (corresponds to $22$ temperatures). Because the overlap between the $3$ temperatures (when $t=e^{5.0}$) is insufficient, there are obviously $3$ peaks in the potential energy distribution, and the peak corresponding to the lowest temperature is very high, indicating a low sampling efficiency of high energy range. While for $t= e^{0.05}$ ($22$ temperatures), a more uniform potential energy distribution is generated. A good choice of overlap factor therefore should ensure sufficient overlap between adjacent temperatures.

\par
\begin{figure}[!htbp]
\begin{center}
\includegraphics[angle=0,width=1.0\textwidth]{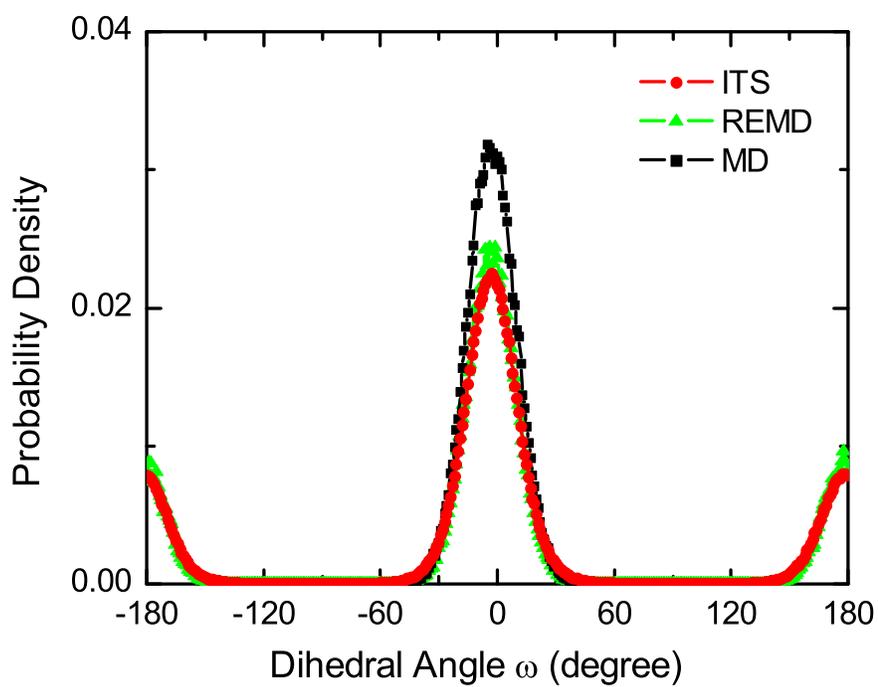}
\end{center}
\caption{The distribution of the dihedral angle $\omega$ in ITS simulation (red circle), REMD simulation (green triangle) and conventional MD simulation (black square). }\label{figure:omega}
\end{figure}

To illustrate the sampling efficiency in configuration space, we compare the dihedral angle ($\omega$) distributions obtained in ITS, REMD and conventional MD simulations, as shown in Fig.~\ref{figure:omega}. Because the free energy barrier is pretty high for trans/cis transition of this dipeptide, in conventional MD simulation, no transition occurs and a unimodal $\omega$ distribution is observed. While in ITS as well as in REMD simulations, bimodal distributions of $\omega$ are observed. The result indicates that both ITS and REMD methods can overcome high free energy barrier and enhance the sampling of configuration space.

\par
\begin{figure}[!htbp]
\begin{center}
\includegraphics[angle=0,width=1.0\textwidth]{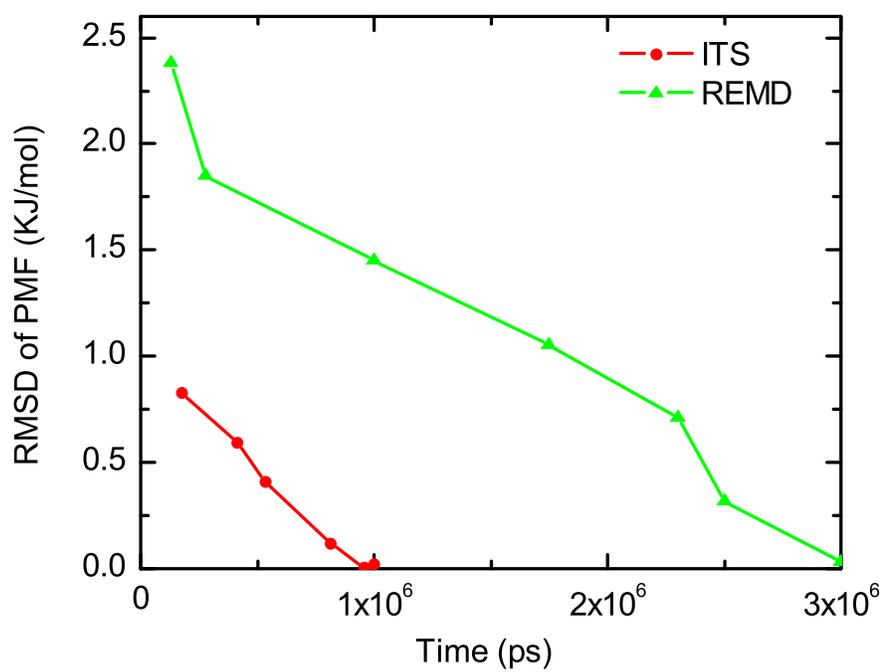}
\end{center}
\caption{The convergence of RMSD of PMF in ITS simulation (red circle) and REMD simulation (green triangle), respectively. It is clear that the RMSD of PMF converges much more quickly in ITS simulation than that in REMD simulation.}
\label{figure:pmfrmsd}
\end{figure}

To further compare the sampling efficiency of ITS and REMD methods, root of mean square derivation (RMSD) of potential of mean force (PMF) along the reaction coordinate ($\omega$) is investigated. PMF along the dihedral angle ($\omega$) is defined as
\begin{equation}
\label{equ:defpmf}
F^{pmf}(\omega) = -k_BT\ln{\langle \rho(\omega) \rangle }\,.
\end{equation}
In Eq.~(\ref{equ:defpmf}), $\langle \rho(\omega) \rangle$ is average density function defined in Eq.~(\ref{equ:defrho})
\begin{equation}
\label{equ:defrho}
\langle \rho(\omega) \rangle = \frac{\int{\delta(\omega'(r)-\omega) e^{-\frac{U(r)}{k_BT} }dr} }{\int{e^{-\frac{U(r)}{k_BT} } dr} }\,,
\end{equation}
which can be calculated through a reweighting scheme using Eq.~(\ref{equ:reweight}). Fig.~\ref{figure:pmfrmsd} shows the time evolution of RMSD of PMF. The RMSD of PMF converges much more quickly in ITS simulation than that in REMD simulation. The result implies that ITS method is more efficient than REMD in sampling configuration and energy space.

\par
Another important advantage of ITS simulation is that it requires less computational resources. It is necessary in REMD simulation to launch several simulations simultaneously, while in ITS simulation, only one trajectory is needed. The computational resources required by ITS is almost the same as conventional MD simulation. In this dipeptide case, the CPU time for REMD simulation is about $263$ hours (eight trajectories in total) and for ITS simulation is about $35$ hours (only one trajectory is needed). For this simplest peptide system, the REMD simulation is nearly $8$ times computational expensive.

\subsection{Coil-globule transition of a flexible single polymer chain}
The transition of a flexible polymer chain from a random-coil conformation to a globular
compact form has been extensively studied~\cite{liting2010coil,liang2000coil,seaton2010coil}. In order to demonstrate the applicability of this ITS method, we apply it to study the coil-globule transition of a flexible single polymer chain in implicit solvent. By calculating the mean square radius of gyration ($\langle Rg^2\rangle$) of the polymer chain at different temperatures, we can get the transition temperature of coil-globule transition.

\par
Conventional reduced units are used in the following. We consider a coarse-grained model of polymer with $100$ beads connected by finite extensible nonlinear elastic (FENE) potential,
\begin{equation}
\label{equ:fene}
U_{FENE}(r) =
\begin{array}{ll}
 -\frac{1}{2}K{r_0}^2\ln({1.0-\frac{r^2}{{r_0}^2}}) + U_{WCA} &\qquad r< r_0\\
 \infty &\qquad r\ge r_0
 \end{array}
 \,,
\end{equation}
in which
\begin{equation}
\label{equ:wca}
 U_{WCA}(r) =  \begin{array}{ll}
 4 \varepsilon_{WCA} \left[ \left( \frac{\sigma_{WCA}}{r} \right)^{12} - \left( \frac{\sigma_{WCA}}{r} \right)^{6} \right] + \varepsilon_{WCA} &\qquad r< 2^{\frac{1}{6}}\sigma_{WCA}\\
    0 &\qquad r \ge 2^{\frac{1}{6}}\sigma_{WCA}  \end{array}\,
\end{equation}
and $r_0$ is the bond extension parameter, $K$ is the attractive force strength, and $r$ is the instantaneous bond length. We set $\sigma_{WCA}$=1.05, $\varepsilon_{WCA}$=1.0,  $r_0$=1.5, and $K$=20. The Lennard-Jones potential in Eq.~(\ref{equ:lj}) is used between non-bonded beads with $\sigma$ =1.0 and $\varepsilon$=1.0. For comparison, we perform both ITS and
MD simulations with the same initial chain configuration and the same parameter set.

\par
\begin{figure}[!htbp]
\begin{center}
\includegraphics[angle=0,width=0.8\textwidth]{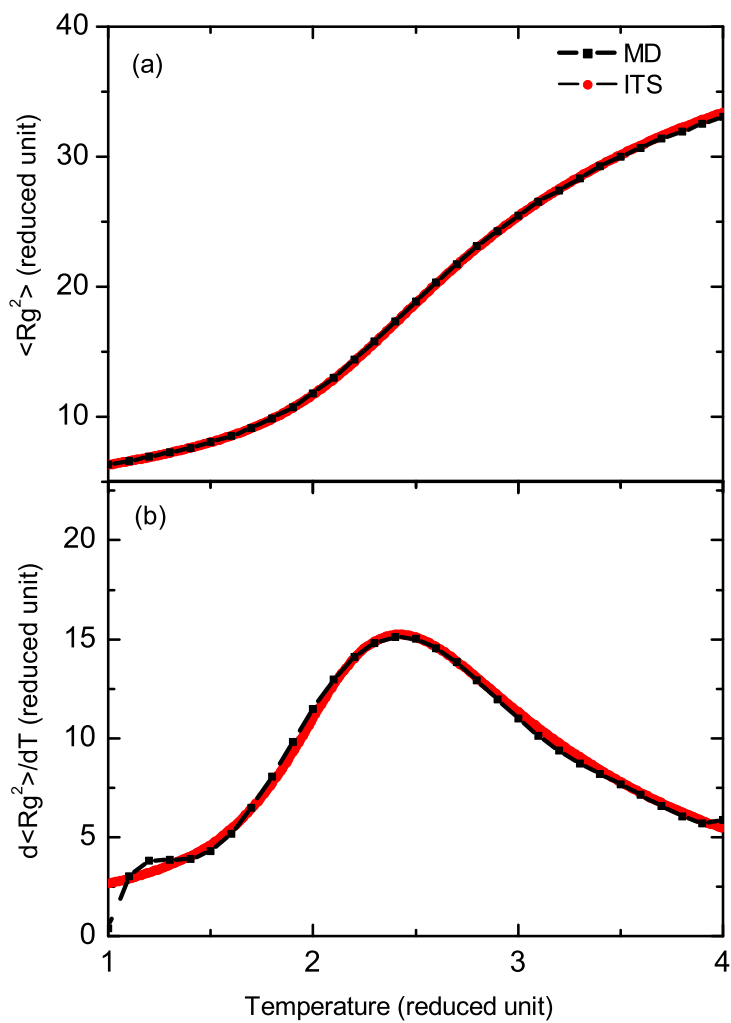}
\end{center}
\caption{The radius of gyration (a) and its derivative (b) versus temperature in ITS simulation (red circle) and MD simulations (black square). The derivative curve for ITS is computed by difference method and the derivative curve for MD is computed by 8-order polynomial fitting.}
\label{figure:rg2t}
\end{figure}

The potential energy curve is obtained by $1.0\times10^6$ steps MD simulation at $8$ temperatures in the range of $1.0\sim 4.5$. We then perform $1.0\times10^9$ steps ITS simulation, for which the overlap factor $t$ is set to $e^{0.5}$ and $18$ temperatures are generated in the range of $1.0\sim 4.35$. By performing one ITS simulation, we can get the $\langle Rg^2\rangle$ at any temperature in the temperature range. By calculating the first order derivative, the transition temperature can be identified, as shown in Fig.~\ref{figure:rg2t} for $\langle Rg^2\rangle$ at $300$ temperatures. As temperature increases, the value of $\langle Rg^2\rangle$ also increases. Because the curve for $\langle Rg^2\rangle$ produced by ITS simulation is smooth with high revolution, we calculate first order and second order derivative directly by difference method.  The peak of the first order derivative of $\langle Rg^2\rangle \sim T$ corresponds to the transition temperature. By calculating the second order derivative, we easily identify the transition temperature as $2.42$ for coil-globule transition of a polymer chain with 100 beads.

\par
For comparison, we also use brute-force canonical MD simulations to study the coil-globule transition of this polymer chain.
We perform $31$  MD simulations at $31$ temperatures spaced $0.1$ in the range of $1.0\sim 4.0$. The length of each simulation is $1.0\times10^9$ timesteps. As shown in Fig.~\ref{figure:rg2t}, the curve of $\langle Rg^2\rangle\sim T$ produced by canonical MD simulations basically overlaps with the one produced by ITS simulation. Because the resolution of the curve produced by MD simulations is low and the noise of data is significant, we employ polynomial fitting method to calculate the derivative. we try different orders of polynomial, and find that the 8-order polynomial can best reproduce the results of ITS simulation. Because the computational cost of ITS simulation is nearly the same as conventional MD simulation and we can get even more accurate data with high resolution in one ITS simulation than massive canonical MD simulations, ITS method is much more efficient than conventional MD on identifying the coil-globule transition temperature for polymer chain. The results indicate that ITS method can be used to study complex polymer systems with high efficiency and accuracy.

\section{Conclusion}\label{sec:conclusion}
In this study, we present a new version of ITS method that provides an easy, quick and robust way to generate suitable parameters. In this method, the only input is potential energy average in the desired range of temperatures. Reasonable values of weighting factors $n_k$ can be obtained directly from potential energy average without iteration, even though the potential energy average is not that accurate. It is also easy to determine the temperature distribution in which there is sufficient overlap between adjacent temperatures by choosing a reasonable overlap factor. This method is very efficient for exploring configuration space and calculating thermodynamic quantities. By running one ITS simulation (i.e., one trajectory), we can sample basically the same configuration space as REMD simulations in the same temperature range. But in ITS method, we do not need to launch tens of parallel simulations simultaneously, so it is extremely suitable to be implemented in GPU version of typical simulation packages for enhanced sampling.

\par
In the method we proposed, to determine the weighting factors in ITS simulation, we use the optimizing condition that when $W(r)$ is dominated by the $k$-th term in the range $U_{k-1}^p < U < U_{k}^p$, the maximum of the potential energy probability density should be in the same range. This optimizing can be easily satisfied even the potential averages are not very accurate, so we could estimate these parameters by short MD simulation trajectories.

\par
Glass transition is fundamental and challenging problem in solid state physics and also an important phenomenon in material science. The debate about whether the glass transition is a thermodynamic phase transition or a dynamic phenomenon has been lasting for decades~\cite{gibbs1958thermo,gordon1976dynamic,elenius2010thermo}. Because this ITS method is a powerful tool to calculate the thermodynamic quantities, we hope this method will contribute to solving the problem of glass transition.

\par
\section*{Acknowledgements}
This work is subsidized by the National Basic Research Program of China (973 Program, 2012CB821500), and supported by  National Science Foundation of China ($21025416$, $50930001$).

\bibliographystyle{model1-num-names}
\bibliography{bibdata}

\newpage
\begin{table}[!htbp]
\begin{center}
\begin{threeparttable}
\caption{Comparison of potential energy per particle (in reduced unit) of Lennard-Jones fluid in ITS simulation and from literature.}\label{table:ljpot}
\begin{tabular}{ccc}
  \hline\hline
  Temperature & ITS results & Literature data~\cite{js1993lj} \\
  \hline
  $1.4$ & $-5.199$ & $-5.199$ \\
  $1.6$ & $-5.045$ & $-5.046$ \\
  $1.8$ & $-4.895$ & $-4.896$ \\
  \hline\hline
\end{tabular}
\end{threeparttable}
\end{center}
\end{table}

\newpage
\begin{table}[!htbp]
\begin{center}
\begin{threeparttable}
\caption{Potential energy averages of ALA-PRO dipeptide calculated from 10 ps and $1~\mu$s trajectories, respectively, and ITS simulation at different temperatures.}\label{table:enerpot}
\begin{tabular}{cccc}\\
  \hline\hline
  Temperature (K) & $10$ ps (KJ/mol)  & $1$ $\mu$s (KJ/mol) & ITS (KJ/mol)\\
  \hline
  $283$ & $-487.83$ & $-491.93$ & $-491.94$\\
  $335$ & $-480.13$ & $-478.51$ & $-478.55$\\
  $400$ & $-459.46$ & $-461.82$ & $-461.88$\\
  $478$ & $-439.46$ & $-441.92$ & $-442.03$\\
  $564$ & $-414.80$ & $-420.18$ & $-420.33$\\
  $680$ & $-369.86$ & $-391.09$ & $-391.35$\\
  $805$ & $-345.94$ & $-359.96$ & $-360.32$\\
  $950$ & $-301.71$ & $-324.01$ & $-324.30$\\
  \hline\hline
\end{tabular}
\end{threeparttable}
\end{center}
\end{table}

\newpage
\begin{table}[!htbp]
\begin{center}
\begin{threeparttable}
\caption{Different overlap factor $t$, number of temperatures, standard deviation of potential energy.}\label{table:t}
\begin{tabular}{c@{\quad}c@{\quad}c@{\quad}c@{\quad}}\\
  \hline\hline
  Overlap  & Number & Standard deviation of potential\\
  factor $t$ & of temperatures &   energy (KJ/mol) \\
  \hline
  $e^{29.8}$ & $2$ & $17.10$ \\
  $e^{5.0}$ & $3$ & $39.71$ \\
  $e^{0.5}$ & $8$ & $53.24$ \\
  $e^{0.05}$ & $22$ & $56.40$ \\
  conventional MD & $1$ & $32.9$ \\
  \hline\hline
\end{tabular}
\end{threeparttable}
\end{center}
\end{table}

\newpage

\newpage

\newpage

\newpage

\newpage

\newpage

\newpage

\newpage

\end{document}